\documentstyle[preprint,aps,rotating]{revtex}
\draft
\begin{document}
\tightenlines
\input epsf
\title{
\begin{flushright}
\small\rm HUB-EP-99/58
\end{flushright}
\bigskip
Quark-antiquark potential with retardation and radiative
contributions and the heavy quarkonium mass
spectra }
\author{ D. Ebert, R. N. Faustov\thanks{On leave of absence 
from Russian Academy of Sciences,
Scientific Council for Cybernetics,
Vavilov Street 40, Moscow 117333, Russia.}
~and V. O. Galkin$^{*}$}

\address{Institut f\"ur Physik, Humboldt--Universit\"at zu Berlin,
Invalidenstr.110, D-10115 Berlin, Germany}

\maketitle
\begin{abstract}
The charmonium and bottomonium mass spectra are calculated with the
systematic account of all relativistic corrections of order $v^2/c^2$ and
the one-loop radiative corrections. Special attention is paid to
the contribution of the retardation effects to the spin-independent
part of the quark-antiquark potential, and a general approach to accounting
for retardation effects in the long-range (confining) part of the
potential is presented. A good fit to available experimental
data on the mass spectra is obtained.  
\end{abstract}

\section{Introduction}
The investigation of the meson properties in the framework of constituent
quark models is an important problem of the elementary particle  physics.
At present a large amount of experimental data on the masses of ground
and excited states of heavy and light mesons has been accumulated \cite{pdg}.
By comparing theoretical predictions with experimental data, one can obtain a 
valuable information on the form of the quark-antiquark interaction potential.
Such information is of great practical interest since at present 
it is not possible
to obtain the $q\bar q$ potential in the whole 
range of distances from the basic
principles of QCD. As it is well known, the growing of the strong coupling
constant with distance makes perturbation 
theory inapplicable at large distances
(in the infrared region). In this region it 
is necessary to account for nonperturbative
effects connected with the complicated structure of the QCD vacuum. 
All this leads to 
a theoretical uncertainty in the $q\bar q$ potential at large and intermediate 
distances. It is just in this region of large and intermediate distances that
most of the basic meson characteristics are formed. This makes it possible
to investigate the low-energy region of strong interaction by studying the
mass spectra and decays of mesons.    

Some recent investigations \cite{sim,bl,gpz} have shown that there could be
also a linear (in radius) correction to the perturbative Coulomb potential
at small distances (in contradiction with OPE predictions). The estimates
of the slope yield that it could be
 of the same order of magnitude as the slope of 
the long-range confining linear potential. It means then that the widely used 
Cornell potential (the sum of the Coulomb and linear confining terms) is 
really a correct one in the static limit both at large and at small distances. 

The relativistic properties of the quark-antiquark interaction potential
play an important role in analysing different static and dynamical 
characteristics of heavy me\-sons. The Lorentz-structure of the confining 
quark-anti\-quark interaction is of particular interest. In the literature 
there is no consent on this item. For a 
long time the scalar confining kernel has been considered to be the
most appropriate one \cite{scal}. The main argument in favour of this
choice is based on the nature of the heavy quark spin-orbit potential.
The scalar potential gives a vanishing long-range magnetic 
contribution, which is in agreement with the flux tube picture
of quark confinement of~\cite{buch}, and allows to get the fine
structure for heavy quarkonia in accord with experimental data. 
However, the calculations of electroweak decay rates of heavy mesons 
with a scalar confining potential alone yield results which are in  worse 
agreement with data than for a vector potential \cite{mb,gf}. 
The  radiative
$M1$-transitions in quarkonia such as e.~g. $J/\psi\to \eta_c
\gamma$ are the most sensitive
to the Lorentz-structure of the confining potential. 
The relativistic corrections for 
these decays arising from vector and scalar potentials have different
signs \cite{mb,gf}. In particular, as it has been 
shown in ref.~\cite{gf}, agreement
with experiments for these decays can be achieved only for a mixture
of vector and scalar potentials. In this context, it is worth remarking,
that the recent study of the $q\bar q$ 
interaction in the Wilson loop approach \cite{bv1} indicates that
it cannot be considered as simply a scalar. Moreover, the found
structure of spin-independent relativistic corrections is not 
compatible with a scalar potential. A similar conclusion
has been obtained in ref.~\cite{ss} on the basis of a Foldy-Wouthuysen 
reduction of the full Coulomb gauge Hamiltonian of QCD. There, the 
Lorentz-structure of the confinement has been found to be of vector
nature. The scalar character of spin splittings in heavy quarkonia
in this approach is dynamically generated through the interaction
with collective gluonic degrees of freedom. Thus we see that while the
spin-dependent structure of ($q\bar q$) interaction is well established
now,  the spin-in\-de\-pen\-dent part is still controversial in the 
literature. 
The uncertainty in the Lorentz-structure of the confining interaction 
complicates the account for retardation corrections since the relativistic
reconstruction of the static confining potential is not unique. In our 
previous paper \cite{pot} we gave some possible prescription of such 
reconstruction which, in particular, provides the fulfilment of the Barchielli,
Brambilla, Prosperi (BBP) relations \cite{bbp} following from the Lorentz
invariance of the Wilson loop. Here we generalize this prescription and 
discuss its connection with the known quark potentials and the implications
for the heavy quarkonium mass spectra. 

The other important point is the inclusion of radiative corrections in the
perturbative part of the quark potential. There have been considerable 
progress in recent years and now the perturbative QCD corrections to the
static potential are known up to two loops \cite{peter,schr} though for
the velocity dependent and spin-dependent potentials only one-loop
corrections are calculated \cite{gupta,ty,ptn}.

The paper is organized as 
follows. In Sec.~II we describe our relativistic quark model. The 
approach to accounting for retardation effects in the $q\bar q$ 
potential in the general case is presented in Sec.~III. The resulting 
heavy quark potential containing both spin-independent and 
spin-dependent parts with the account of one-loop radiative
corrections is given in Sec.~IV. We use this potential for the calculations
of the heavy quarkonium mass spectra in Sec.~V.
Section~VI contains our conclusions and discussion of the results.

\section{Relativistic quark model}  
\label{rqm}

In the quasipotential approach a meson is described by the wave
function of the bound quark-antiquark state, which satisfies the
quasipotential equation \cite{3} of the Schr\"odinger type \cite{4}
\begin{equation}
\label{quas}
{\left(\frac{b^2(M)}{2\mu_{R}}-\frac{{\bf
p}^2}{2\mu_{R}}\right)\Psi_{M}({\bf p})} =\int\frac{d^3 q}{(2\pi)^3}
 V({\bf p,q};M)\Psi_{M}({\bf q}),
\end{equation}
where the relativistic reduced mass is
\begin{equation}
\mu_{R}=\frac{E_aE_b}{E_a+E_b}=\frac{M^4-(m^2_a-m^2_b)^2}{4M^3},
\end{equation}
and $E_a$, $E_b$ are given by
\begin{equation}
\label{ee}
E_a=\frac{M^2-m_b^2+m_a^2}{2M}, \quad E_b=\frac{M^2-m_a^2+m_b^2}{2M}.
\end{equation}
Here $M=E_a+E_b$ is the meson mass, $m_{a,b}$ are the masses of light
and heavy quarks, and ${\bf p}$ is their relative momentum.  
In the centre of mass system the relative momentum squared on mass shell 
reads
\begin{equation}
{b^2(M) }
=\frac{[M^2-(m_a+m_b)^2][M^2-(m_a-m_b)^2]}{4M^2}.
\end{equation}

The kernel 
$V({\bf p,q};M)$ in Eq.~(\ref{quas}) is the quasipotential operator of
the quark-antiquark interaction. It is constructed with the help of the
off-mass-shell scattering amplitude, projected onto the positive
energy states. 
Constructing the quasipotential of the quark-antiquark interaction 
we have assumed that the effective
interaction is the sum of the usual one-gluon exchange term with the mixture
of long-range vector and scalar linear confining potentials, where
the vector confining potential
contains the Pauli interaction. The quasipotential is then defined by
\cite{mass}  
\begin{eqnarray}
\label{qpot}
V({\bf p,q};M)&=&\bar{u}_a(p)
\bar{u}_b(-p)\Bigg\{\frac{4}{3}\alpha_sD_{ \mu\nu}({\bf
k})\gamma_a^{\mu}\gamma_b^{\nu}\cr
& & +V_V({\bf k})\Gamma_a^{\mu}
\Gamma_{b;\mu}+V_S({\bf
k})\Bigg\}u_a(q)u_b(-q),
\end{eqnarray}
where $\alpha_S$ is the QCD coupling constant, $D_{\mu\nu}$ is the
gluon propagator in the Coulomb gauge
\begin{equation}
D^{00}({\bf k})=-\frac{4\pi}{{\bf k}^2}, \quad D^{ij}({\bf k})=
-\frac{4\pi}{k^2}\left(\delta^{ij}-\frac{k^ik^j}{{\bf k}^2}\right),
\quad D^{0i}=D^{i0}=0,
\end{equation}
and ${\bf k=p-q}$; $\gamma_{\mu}$ and $u(p)$ are 
the Dirac matrices and spinors
\begin{equation}
\label{spinor}
u^\lambda({p})=\sqrt{\frac{\epsilon(p)+m}{2\epsilon(p)}}
{1\choose \frac{\bbox{\sigma p}}{\epsilon(p)+m}}\chi^\lambda,
\end{equation}
with $\epsilon(p)=\sqrt{p^2+m^2}$.
The effective long-range vector vertex is
given by
\begin{equation}
\label{kappa}
\Gamma_{\mu}({\bf k})=\gamma_{\mu}+
\frac{i\kappa}{2m}\sigma_{\mu\nu}k^{\nu},
\end{equation}
where $\kappa$ is the Pauli interaction constant characterizing the
anomalous chromomagnetic moment of quarks. Vector and
scalar confining potentials in the nonrelativistic limit reduce to
\begin{eqnarray}
\label{vlin}
V_V(r)&=&(1-\varepsilon)Ar+B,\nonumber\\
V_S(r)& =&\varepsilon Ar,
\end{eqnarray}
reproducing 
\begin{equation}
\label{nr}
V_{\rm conf}(r)=V_S(r)+V_V(r)=Ar+B,
\end{equation}
where $\varepsilon$ is the mixing coefficient. 

The expression for the quasipotential for the heavy quarkonia,
expanded in $v^2/c^2$ without retardation corrections to the
 confining potential,
can be found in Ref.~\cite{mass}. The 
structure of the spin-dependent interaction is in agreement with
the parameterization of Eichten and Feinberg \cite{ef}.  
All the parameters of
our model like quark masses, parameters of the linear confining potential
$A$ and $B$, mixing coefficient $\varepsilon$ and anomalous
chromomagnetic quark moment $\kappa$ are fixed from the analysis of
heavy quarkonium masses (see below Sec.~V) and radiative decays. 
The quark masses
$m_b=4.88$ GeV, $m_c=1.55$ GeV and
the parameters of the linear potential $A=0.18$ GeV$^2$ and $B=-0.16$ GeV
have usual values of quark models.  The value of the mixing
coefficient of vector and scalar confining potentials $\varepsilon=-1$
has been determined from the consideration of the heavy quark expansion
for the semileptonic $B\to D$ decays
\cite{fg} and charmonium radiative decays \cite{gf}.
Finally, the universal Pauli interaction constant $\kappa=-1$ has been
fixed from the analysis of the fine splitting of heavy quarkonia ${
}^3P_J$- states \cite{mass}. Note that the 
long-range  magnetic contribution to the potential in our model
is proportional to $(1+\kappa)$ and thus vanishes for the 
chosen value of $\kappa=-1$. In  the present paper we will include 
into consideration the retardation corrections as well as one-loop  
radiative corrections. 

\section{General approach to accounting for retardation effects in the 
{$\bbox{q\bar q}$} potential}

For the one-gluon exchange part of the $q\bar q$ potential
 it is quite easy to isolate
the retardation contribution. Indeed due to the 
vector current conservation (gauge
invariance) we have the well-known relation on the mass shell
\begin{eqnarray}
\label{cpr}
&&\frac{1}{k^2}\bar u_{a}({\bf p})\bar u_b(-{\bf p})\gamma^\mu_a\gamma_{b\mu}
u_a({\bf q})u_b({-\bf q})\cr
&& \qquad=-\bar u_{a}({\bf p})\bar u_b(-{\bf p})\left\{\frac{\gamma_a^0
\gamma_b^0}{{\bf k}^2} +\frac{1}{k^2}
\left[{\bbox{\gamma}_a\cdot \bbox{\gamma}_b}
-\frac{(\bbox{\gamma}_a\cdot {\bf k})(\bbox{\gamma}_b
\cdot {\bf k})}{{\bf k}^2}\right]\right\}
 u_a({\bf q})u_b({-\bf q}), \\
&& k^2=k_0^2-{\bf k}^2; \quad k_0=\epsilon_a({\bf p})-\epsilon_a({\bf q})=
\epsilon_b({\bf q})-\epsilon_b({\bf p}); 
\quad {\bf k}={\bf p}-{\bf q}.\nonumber
\end{eqnarray}
The left-hand side and the right-hand side of this 
relation are easily recognized to be 
in the Feynman gauge and the Coulomb gauge, respectively. Now,
 if the nonrelativistic
expansion in $p^2/m^2$ is applicable, 
we can immediately extract the retardation
contribution. Namely we expand the left-hand side of 
eq.~(\ref{cpr}) in $k^2_0/{\bf k}^2$:
$$ \frac{1}{k_0^2-{\bf k}^2}\cong -\frac{1}{{\bf k}^2}-
\frac{k_0^2}{{\bf k}^4}$$
and get with needed accuracy \cite{ab}
\begin{equation}
\label{rel}
-\bar u_{a}({\bf p})\bar u_b(-{\bf p})\left[\frac{\gamma_a^0
\gamma_b^0}{{\bf k}^2}\left(1+\frac{k_0^2}{{\bf k}^2}\right)
-\frac{\bbox{\gamma}_a\cdot \bbox{\gamma}_b}{{\bf k}^2}\right]
 u_a({\bf q})u_b({-\bf q}).
\end{equation}
In the right-hand side of eq.~(\ref{cpr}) one should use the identity following from the 
Dirac equation
\begin{eqnarray}
&&\bar u_{a}({\bf p})\bar u_b(-{\bf p})
(\bbox{\gamma}_a\cdot {\bf k})(\bbox{\gamma}_b\cdot
{\bf k})u_a({\bf q})u_b({-\bf q})\cr
&&\qquad=\bar u_{a}({\bf p})\bar u_b(-{\bf p})\gamma_a^0
\gamma_b^0u_a({\bf q})u_b({-\bf q})(\epsilon_a({\bf p})-\epsilon_a({\bf q}))
(\epsilon_b({\bf q})-\epsilon_b({\bf p})).\nonumber
\end{eqnarray}
After defining $k_0^2$ as a symmetrized product \cite{ab,gromes}
\begin{equation}
\label{k0}
k_0^2=(\epsilon_a({\bf p})-\epsilon_a({\bf q}))(\epsilon_b({\bf q})-\epsilon_b({\bf p}))
\end{equation}
and dropping $k_0^2$ in the denominator we obtain the expression which is 
identical to eq.~(\ref{rel}). In this way we obtain the well-known Breit Hamiltonian
(the same as in QED \cite{ab}) if we further expand eq.~(\ref{k0}) in $p^2/m^2$
\begin{equation}
\label{k02}
k_0^2\cong-\frac{({\bf p}^2-{\bf q}^2)^2}{4m_am_b}.
\end{equation}
This treatment allows also for the correct Dirac limit 
in which the retardation contribution vanishes when one of the particles
becomes infinitely heavy  \cite{om}.

For the confining part of the $q\bar q$ potential the retardation contribution is much
more indefinite. It is a consequence of our poor knowledge of the confining potential
especially in what concerns its relativistic properties: the Lorentz structure
(scalar, vector, etc.) and the dependence on the covariant variables such as 
$k^2=k_0^2-{\bf k}^2$. Nevertheless we can perform some general considerations and then apply them to a particular case of the linearly rising potential. To this end we
note that for any nonrelativistic potential $V(-{\bf k}^2)$ the simplest relativistic
generalization is to replace it by $V(k_0^2-{\bf k}^2)$.

In the case of the Lorentz-vector confining potential we can use the same approach
as before even with more general vertices containing the Pauli terms, since the
mass-shell vector currents are conserved here as well. It is possible to introduce
alongside with the ``diagonal gauge" the so-called ``instantaneous gauge'' 
\cite{ch} which 
is the generalization of the Coulomb gauge. The relation analogous to 
eq.~(\ref{cpr}) now looks like (up to the terms of order of $p^2/m^2$)
\begin{eqnarray}
\label{vpr}
&&V_V(k_0^2-{\bf k}^2)\bar u_{a}({\bf p})\bar u_b(-{\bf p})
\Gamma^\mu_a\Gamma_{b\mu}
u_a({\bf q})u_b({-\bf q})
=\bar u_{a}({\bf p})\bar u_b(-{\bf p})\biggl\{V_V(-{\bf k}^2){\Gamma_a^0
\Gamma_b^0}\cr
&& \qquad -\left[V_V(-{\bf k}^2){\bbox{\Gamma}_a\cdot \bbox{\Gamma}_b}
+V'_V(-{\bf k}^2)(\bbox{\Gamma}_a \cdot
{\bf k})(\bbox{\Gamma}_b\cdot{\bf k})\right]\biggr\}
 u_a({\bf q})u_b({-\bf q}),
\end{eqnarray}
where
$$V_V(k_0^2-{\bf k}^2)\cong V_V(-{\bf k}^2)+k_0^2V'_V(-{\bf k}^2)$$
and as in the case of the one-gluon exchange above we put
\begin{equation}
\label{k0s}
k_0^2=(\epsilon_a({\bf p})-\epsilon_a({\bf q}))(\epsilon_b({\bf q})-\epsilon_b({\bf p}))
\cong-\frac{({\bf p}^2-{\bf q}^2)^2}{4m_am_b}
\end{equation}
again with the correct Dirac limit.

For the case of the Lorentz-scalar potential we can make the same expansion 
in $k_0^2$,
which yields
\begin{equation}
\label{vse}
V_S(k_0^2-{\bf k}^2)\cong V_S(-{\bf k}^2)+k_0^2V'_S(-{\bf k}^2).
\end{equation}
But in this case we have no reasons to fix $k_0^2$ in the only way (\ref{k0}).
The other possibility is to take a half sum instead of a symmetrized product,
namely to set (see e.~g. \cite{gromes,om})
\begin{equation}
\label{k0hs}
k_0^2=\frac12\left[(\epsilon_a({\bf p})-\epsilon_a({\bf q}))^2+
(\epsilon_b({\bf q})-\epsilon_b({\bf p}))^2\right]\cong
\frac18({\bf p}^2-{\bf q})^2\left(\frac{1}{m_a^2}+\frac{1}{m_b^2}\right).
\end{equation}
The Dirac limit is not fulfilled by this choice, but this 
cannot serve as a decisive
argument. Thus the most general expression for the energy transfer squared,
which incorporates both possibilities (\ref{k0s}) and (\ref{k0hs}) has the form
\begin{equation}
\label{k00}
k_0^2=\lambda(\epsilon_a({\bf p})-\epsilon_a({\bf q}))
(\epsilon_b({\bf q})-\epsilon_b({\bf p}))
+(1-\lambda)\frac12\left[(\epsilon_a({\bf p})-\epsilon_a({\bf q}))^2+
(\epsilon_b({\bf q})-\epsilon_b({\bf p}))^2\right],
\end{equation}
where $\lambda$ is the mixing parameter.

After making expansion in $p^2/m^2$ we obtain
\begin{eqnarray}
\label{k02s}
k_0^2&\cong&-\lambda\frac{({\bf p}^2-{\bf q}^2)^2}{4m_am_b}
+(1-\lambda)\frac18({\bf p}^2-{\bf q})^2\left(\frac{1}{m_a^2}+\frac{1}{m_b^2}\right)\cr
&=&\frac18\left[(1-\lambda)\left(\frac{1}{m_a^2}+\frac{1}{m_b^2}\right)
-\frac{2\lambda}{m_am_b}\right]\left[({\bf k\cdot p})^2
+2({\bf k\cdot p})({\bf k\cdot q})+({\bf k\cdot q})^2\right].
\end{eqnarray}
Thus as expected $k_0^2\sim O(p^2/m^2)\ll 1$. Then the Fourier transform of the
potential
$$V(k_0^2-{\bf k}^2)\cong V(-{\bf k}^2)+k_0^2V'(-{\bf k})^2$$
with $k_0^2$ given by  eq.~(\ref{k02s}) can be represented as follows \cite{om}
\begin{eqnarray}
\label{ret}
\int\frac{d^3k}{(2\pi)^3}V(k_0^2-{\bf k}^2)e^{i{\bf k\cdot r}}&=&V(r)+\frac14\Biggl[
(1-\lambda)\left(\frac{1}{m_a^2}+\frac{1}{m_b^2}\right)-\frac{2\lambda}{m_am_b}
\Biggr]\cr
&&\times \left\{V(r){\bf p}^2+V'(r)\frac1r({\bf p\cdot r})^2\right\}_W,
\end{eqnarray}
where $\{\dots\}_W$ denotes the Weyl ordering of operators and
\begin{equation}
V(r)=\int\frac{d^3k}{(2\pi)^3}V(-{\bf k}^2)e^{i{\bf k\cdot r}}.
\end{equation}
In the case of the one-gluon exchange potential we had $\lambda=1$,
\begin{equation}
\label{cnr}
V_C(-{\bf k}^2)=-\frac43\frac{4\pi\alpha_s}{{\bf k}^2};
\quad V_C(r)=-\frac43\frac{\alpha_s}{r}.
\end{equation}

As for the confining potential we assume it to be a mixture of scalar and vector parts.
In the nonrelativistic limit we adopt the linearly rising potential
\begin{equation}
V_0(r)=Ar; \qquad V_0(-{\bf k}^2)=-\frac{8\pi A}{({\bf k}^2)^2},
\end{equation}
which we split into scalar and vector parts by introducing the mixing parameter
$\varepsilon$. The possible constant term in $V_0$ has been discussed in \cite{pot}.
\begin{equation}
\label{confnr}
V_0=V_S+V_V; \quad V_S=\varepsilon V_0; \quad V_V=(1-\varepsilon) V_0.
\end{equation}
Hence the retardation contribution (\ref{ret}) from scalar and vector potentials has 
the form
\begin{equation}
\label{cret}
\frac14\Biggl[
(1-\lambda_{S,V})\left(\frac{1}{m_a^2}+\frac{1}{m_b^2}\right)\
-\frac{2\lambda_{S,V}}{m_am_b}
\Biggr]\left\{V_{S,V}(r){\bf p}^2+V_{S,V}'(r)\frac1r({\bf p\cdot r})^2\right\}_W,
\end{equation}
where we use the general Ansatz (\ref{k00}), (\ref{k02s}) for both the scalar and
vector potentials for the sake of completeness.

The other spin-independent corrections in our model had been calculated earlier 
\cite{mass,pot}
\begin{equation}
\frac18(1+2\kappa)\left(\frac{1}{m_a^2}+
\frac{1}{m_b^2}\right)\Delta V_V(r)
+\frac{1}{m_am_b}\left\{V_V(r){\bf p}^2\right\}_W-\frac{1}{2}\left(\frac{1}{m_a^2}
+\frac{1}{m_b^2}\right)
\left\{V_S(r){\bf p}^2\right\}_W. 
\end{equation}

Adding to the above expression the retardation contributions (\ref{cret}) and the 
nonrelativistic parts (\ref{cnr}) and (\ref{confnr}) we obtain the complete 
spin-independent $q\bar q$ potential:
\begin{equation}
\label{spind}
V_{\rm SI}(r)=V_C(r)+V_0(r) + V_{\rm VD}(r)+\frac18\left(\frac{1}{m_a^2}+
\frac{1}{m_b^2}\right) \Delta\big[V_C(r) 
 +(1+2\kappa)V_V\big],
\end{equation}
where the velocity-dependent part
\begin{eqnarray}
\label{vd}
V_{\rm VD}(r)&=& V_{\rm VD}^C(r)+V_{\rm VD}^V(r)+V_{\rm VD}^S(r),\\
V_{\rm VD}^C(r)&=&\frac{1}{2m_am_b}\left\{V_C(r)\left[{\bf p}^2
+\frac{({\bf p\cdot r})^2}{r^2}\right]\right\}_W 
=\frac{1}{2m_am_b}\left\{-\frac43\frac{\alpha_s}{r}
\left[{\bf p}^2+\frac{({\bf p\cdot r})^2}{r^2}\right]\right\}_W, \cr
V_{\rm VD}^V(r)&=&\frac{1}{m_am_b}\left\{V_V(r){\bf p}^2\right\}_W+
\frac14\Biggl[
(1-\lambda_V)\left(\frac{1}{m_a^2}+\frac{1}{m_b^2}\right)-\frac{2\lambda_V}{m_am_b}
\Biggr]\cr
&&\times \left\{V_V(r){\bf p}^2+V_V'(r)\frac{({\bf p\cdot r})^2}{r}\right\}_W=
(1-\varepsilon)\frac{(1-\lambda_V)}{4}\left(\frac{1}{m_a^2}+\frac{1}{m_b^2}\right)\cr
&&\times\left\{Ar\left[{\bf p}^2+\frac{({\bf p\cdot r})^2}{r^2}\right]
\right\}_W+\frac{(1-\varepsilon)}{m_am_b}
\left\{Ar\left[\left(1-\frac{\lambda_V}{2}
\right){\bf p}^2-\frac{\lambda_V}{2}\frac{({\bf p\cdot r})^2}{r^2}\right]
\right\}_W ,\cr 
V_{\rm VD}^S(r)&=&\frac{1}{2}\left(\frac{1}{m_a^2}
+\frac{1}{m_b^2}\right)\left\{V_V(r){\bf p}^2\right\}_W+
\frac14\Biggl[
(1-\lambda_S)\left(\frac{1}{m_a^2}+\frac{1}{m_b^2}\right)-\frac{2\lambda_S}{m_am_b}
\Biggr]\cr
&&\times \left\{V_V(r){\bf p}^2+V_V'(r)\frac{({\bf p\cdot r})^2}{r}\right\}_W\cr
&=&
-\frac{\varepsilon}{4}\left(\frac{1}{m_a^2}+\frac{1}{m_b^2}\right)
\left\{Ar\left[(1+\lambda_S){\bf p}^2+
(\lambda_S-1)\frac{({\bf p\cdot r})^2}{r^2}\right]
\right\}_W\cr
&&-\frac{\varepsilon\lambda_S}{2m_am_b}
\left\{Ar\left[{\bf p}^2+\frac{({\bf p\cdot r})^2}{r^2}\right]
\right\}_W. \nonumber 
\end{eqnarray}
Making the natural decomposition
\begin{equation}
\label{vdrel}
V_{\rm VD}(r)=\frac{1}{m_am_b}\left\{{\bf p}^2V_{bc}(r)+\frac{({\bf p
\cdot r})^2}{r^2}V_c(r)\right\}_W 
+\left(\frac{1}{m_a^2}+\frac{1}{m_b^2}\right)\left\{{\bf p}^2 V_{de}
(r) -\frac{({\bf p\cdot r})^2}{r^2}V_e(r)\right\}_W
\end{equation}
we obtain from eqs.~(\ref{vd})
\begin{eqnarray}
\label{coef}
&&V_{bc}(r)=-\frac{2\alpha_s}{3r}+\left[(1-\varepsilon)\left(1-\frac{\lambda_V}{2}
\right)-\varepsilon\frac{\lambda_S}{2}\right]Ar,\cr
&&V_c(r)=-\frac{2\alpha_s}{3r}-\left[(1-\varepsilon)\frac{\lambda_V}{2}
+\varepsilon\frac{\lambda_S}{2}\right]Ar,\cr
&&V_{de}(r)=\frac14\left[(1-\varepsilon)\left(1-\lambda_V
\right)-\varepsilon(1+\lambda_S)\right]Ar,\cr
&& V_e(r)=-\frac14\left[(1-\varepsilon)\left(1-\lambda_V
\right)+\varepsilon(1-\lambda_S)\right]Ar.
\end{eqnarray}
The following simple relations hold:
\begin{equation}
\label{srl}
V_{bc}-V_c=(1-\varepsilon)Ar; \quad V_{de}+V_e=-\frac{\varepsilon}{2}Ar.
\end{equation}
The exact BBP relations \cite{bbp} (see also \cite{ck})
in our notations look like
\begin{eqnarray}
\label{re}
&&V_{de}-\frac12 V_{bc}+\frac14(V_C+V_0)=0, \cr
&&V_e+\frac12 V_c+\frac{r}{4}\frac{{\rm d}(V_C+ V_0)}{{\rm d} r}=0
\end{eqnarray}
(in the original version $V_{bc}\equiv-V_b-\frac13 V_c$ and 
$V_{de}\equiv V_d+\frac13V_e$).

The functions (\ref{coef}) identically satisfy the BBP relations (\ref{re})
independently of values of the parameters $\varepsilon$, $\lambda_V$, $\lambda_S$
but only with the account of retardation corrections.

In our model \cite{mass,pot} we have $\varepsilon=-1$ and $\lambda_V=1$, if we assume
further that $\lambda_S=1$ \cite{pot} then
we get
\begin{eqnarray}
&& V_{bc}(r)=-\frac{2\alpha_s}{3r}+\frac32Ar; \quad V_c(r)=-\frac{2\alpha_s}{3r}
-\frac12Ar;\cr
&& V_{de}(r)=\frac12Ar; \quad V_e(r)=0.
\end{eqnarray}
Our expressions (\ref{spind}) and (\ref{vd}) for purely vector ($\varepsilon
=0$) and purely scalar ($\varepsilon=1$) interactions and for $\kappa=0$,
$\lambda_S=\lambda_V=1$ coincide with those of Ref.~\cite{om}.

In the minimal area low (MAL) and flux tube models \cite{bv}
\begin{eqnarray}
&& V_{bc}(r)=-\frac{2\alpha_s}{3r}+\frac16 Ar;
\quad V_c(r)=-\frac{2\alpha_s}{3r}-\frac16 Ar;\cr
&& V_{de}(r)=-\frac16 Ar; \quad V_e=-\frac16 Ar.
\end{eqnarray}
To obtain these expressions one should set in relations (\ref{coef}), (\ref{srl})
\begin{equation}
\varepsilon=\frac23; \qquad \lambda_V+2\lambda_S=1.
\end{equation}
Thus one gets a family of values for $\lambda_V$ and $\lambda_S$. The most natural 
choice reads as
\begin{equation}
\lambda_V=1, \qquad \lambda_S=0,
\end{equation}
which resembles the Gromes proposal \cite{gromes}: the symmetrized product for
the vector potential and the half sum for the scalar potential. But still the Dirac limit is not 
fulfilled in this case.

Expression (\ref{spind}) for $V_{\rm SI}$ contains also the term with the Laplacian:
\begin{equation}
\label{lapl}
\frac18\left(\frac{1}{m_a^2}+\frac{1}{m_b^2}\right) \Delta\big[V_C(r) 
 +(1+2\kappa)V_V(r)\big].
\end{equation}
In the MAL and some other models these terms look like \cite{bv}
\begin{equation}
\frac18\left(\frac{1}{m_a^2}+\frac{1}{m_b^2}\right) \Delta\big[V_C(r)+V_0(r) 
 +V_a(r)\big]
\end{equation}
and usually it is adopted that 
\begin{equation}
\label{mal}
\Delta V_a(r)=0.
\end{equation}
Lattice simulations \cite{bali} suggest that 
\begin{equation}
\label{lat}
\Delta V_a^L(r)= c-\frac{b}{r}, \quad b\cong 0.8 {\rm GeV}^2.
\end{equation}
In our model expression (\ref{lapl}) can be recast as follows
\begin{eqnarray}
&&\frac18\left(\frac{1}{m_a^2}+\frac{1}{m_b^2}\right) \Delta\big[V_C(r) 
 +V_0(r)+\tilde V_a(r)\big],\cr
&&\tilde V_a(r)=(1+2\kappa)(1-\varepsilon)V_0(r)-V_0(r)
\end{eqnarray}
and for the adopted values $\varepsilon=-1$, $\kappa=-1$
\begin{equation}
\Delta\tilde V_a(r)=-3\Delta(Ar)=-6\frac{A}{r}, \quad 6A\cong 1.1 {\rm GeV}^2,
\end{equation}
which is close to the lattice result (\ref{lat}) but differs from the 
suggestion (\ref{mal}).

 \section{Heavy quark-antiquark potential with the account of retardation 
effects and 
one loop radiative corrections}

At present the static quark-antiquark potential in QCD is known to two loops 
\cite{peter,schr}. However the velocity dependent and 
spin-dependent parts are known only
to the one-loop order \cite{gupta,ty}. Thus we limit our analysis to one-loop
radiative corrections. The resulting heavy quark-antiquark potential can be 
presented in the form of a sum of  spin-independent and spin-dependent
parts. For the spin-independent part 
using the relations (\ref{spind}), (\ref{vd})
with $\lambda_V=1$ and including one-loop radiative 
corrections in $\overline {MS}$
renormalization scheme \cite{gupta,ty} we get 
\begin{eqnarray}
\label{sipot}
V_{\rm SI}(r)&=&-\frac43\frac{\bar \alpha_V(\mu^2)}{r} +Ar+B -\frac43\frac{\beta_0
\alpha_s^2(\mu^2)}{2\pi}\frac{\ln(\mu r)}{r} \cr
&& +\frac18\left(\frac{1}{m_a^2}+\frac{1}{m_b^2}\right) \Delta\left[ -\frac43\frac{\bar 
\alpha_V(\mu^2)}{r} -\frac43\frac{\beta_0\alpha_s^2(\mu^2)}{2\pi}\frac{\ln(\mu r)}{r}
 +(1-\varepsilon)(1+2\kappa)Ar\right]\cr
&&+\frac{1}{2m_am_b}\Biggl(\left\{-\frac43\frac{\bar\alpha_V}{r}
\left[{\bf p}^2+\frac{({\bf p\cdot r})^2}{r^2}\right]\right\}_W\cr
&&-\frac43\frac{\beta_0\alpha_s^2(\mu^2)}{2\pi}\left\{{\bf p}^2\frac{\ln(\mu r)}{r}
+\frac{({\bf p\cdot 
r})^2}{r^2}\left(\frac{\ln(\mu r)}{r}-\frac1r\right)\right\}_W\Biggr)\cr
&& +\left[\frac{1-\varepsilon}{2m_am_b}-\frac{\varepsilon}{4}
\left(\frac{1}{m_a^2}+\frac{1}{m_b^2}\right)\right]\left\{Ar\left[{\bf p}^2
-\frac{({\bf p\cdot r})^2}{r^2}\right]\right\}_W\cr
&&-\frac{\varepsilon\lambda_S}{2}\left[\frac12
\left(\frac{1}{m_a^2}+\frac{1}{m_b^2}\right)+\frac{1}{m_am_b}\right]
\left\{Ar\left[{\bf p}^2+\frac{({\bf p\cdot r})^2}{r^2}\right]\right\}_W\cr
&&+\left[\frac{1}{4}\left(\frac{1}{m_a^2}+\frac{1}{m_b^2}\right)+
\frac{1}{m_am_b}\right]B{\bf p}^2,
\end{eqnarray}
where
\begin{eqnarray}
\bar\alpha_V(\mu^2)&=&\alpha_s(\mu^2)\left[1+\left(\frac{a_1}{4}
+\frac{\gamma_E\beta_0}{2}\right)\frac{\alpha_s(\mu^2)}{\pi}\right],\\
a_1&=&\frac{31}{3}-\frac{10}{9}n_f,\cr
\beta_0&=&11-\frac23n_f.\nonumber
\end{eqnarray}
Here $n_f$ is a number of flavours and $\mu$ is a renormalization scale.

For the dependence of the QCD coupling constant $\alpha_s(\mu^2)$ 
on the renormalization point $\mu^2$ we use the leading order
result
\begin{equation}
\label{alpha}
\alpha_s(\mu^2)=\frac{4\pi}{\beta_0\ln(\mu^2/\Lambda^2)}.
\end{equation}

Comparing this expression for $V_{\rm SI}$ with the
decomposition (\ref{vdrel}) we find
\begin{eqnarray}
\label{potcoef}
V_{bc}(r)&=&-\frac23\frac{\bar \alpha_V(\mu^2)}{r}  -\frac23\frac{\beta_0
\alpha_s^2(\mu^2)}{2\pi}\frac{\ln(\mu r)}{r} +\left(\frac{1-\varepsilon}{2}
-\frac{\varepsilon\lambda_S}{2}\right)Ar+B,\cr
V_c(r)&=&-\frac23\frac{\bar \alpha_V(\mu^2)}{r}  -\frac23\frac{\beta_0
\alpha_s^2(\mu^2)}{2\pi}\left[\frac{\ln(\mu r)}{r}-\frac1r\right]
-\left(\frac{1-\varepsilon}{2}+\frac{\varepsilon\lambda_S}{2}\right)Ar,\cr
V_{de}(r)&=& -\frac{\varepsilon}{4}(1+\lambda_S)Ar+B,\cr
V_e(r)&=& -\frac{\varepsilon}{4}(1-\lambda_S)Ar.
\end{eqnarray}
It is easy to check that the BBP relations are exactly satisfied.

The spin-dependent part of the quark-antiquark potential for equal
quark masses ($m_a=m_b=m$)  with the inclusion
of radiative corrections \cite{gupta,ptn} can be presented in our model 
\cite{mass} as follows:
\begin{eqnarray}
\label{vsd}
V_{\rm SD}&=& a\ {\bf L}\cdot{\bf S}+b\left[\frac{3}{r^2}({\bf S}_a\cdot {\bf r})
({\bf S}_b\cdot {\bf r})-({\bf S}_a\cdot {\bf S}_b)\right] +c\ {\bf S}_a\cdot {\bf S}_b, \\
\label{a}
a&=& \frac{1}{2m^2}\Biggl\{\frac{4\alpha_s(\mu^2)}{r^3}\Biggl(1+
\frac{\alpha_s(\mu^2)}{\pi}\Biggl[\frac{1}{18}n_f-\frac{1}{36}+\gamma_E\left(
\frac{\beta_0}{2}-2\right)+\frac{\beta_0}{2}\ln\frac{\mu}{m}\cr
&&+\left(\frac{\beta_0}{2}-2\right)\ln(mr)\Biggr]\Biggr)
-\frac{A}{r}+4(1+\kappa)(1-\varepsilon)\frac{A}{r}\Biggr\}\\
\label{b}
b&=& \frac{1}{3m^2}\Biggl\{\frac{4\alpha_s(\mu^2)}{r^3}\Biggl(1+
\frac{\alpha_s(\mu^2)}{\pi}\Biggl[\frac{1}{6}n_f+\frac{25}{12}+\gamma_E\left(
\frac{\beta_0}{2}-3\right)+\frac{\beta_0}{2}\ln\frac{\mu}{m}\cr
&&+\left(\frac{\beta_0}{2}-3\right)\ln(mr)\Biggr]\Biggr)
+(1+\kappa)^2(1-\varepsilon)\frac{A}{r}\Biggr\}\\
\label{c}
c&=& \frac{4}{3m^2}\Biggl\{\frac{8\pi\alpha_s(\mu^2)}{3}\Biggl(\left[1+
\frac{\alpha_s(\mu^2)}{\pi}\left(\frac{23}{12}-\frac{5}{18}n_f-\frac34\ln2\right)
\right]\delta^3(r)\cr
&&+\frac{\alpha_s(\mu^2)}{\pi}\left[-
\frac{\beta_0}{8\pi}\nabla^2\left(\frac{\ln({\mu}/{m})}{r}\right)
+\frac{1}{\pi}\left(\frac{1}{12}n_f-\frac{1}{16}\right)\nabla^2\left(
\frac{\ln(mr)+\gamma_E}{r}\right)\right]\Biggr)\cr
&&+(1+\kappa)^2(1-\varepsilon)\frac{A}{r}\Biggr\},
\end{eqnarray}
where ${\bf L}$ is the orbital momentum and ${\bf S}_{a,b}$, ${\bf S}={\bf S}_a+
{\bf S}_b$ are the spin momenta.

The correct description of the fine structure of the heavy quarkonium
mass spectrum requires the vanishing of the vector confinement contribution.
This can be achieved by setting $1+\kappa=0$, i.e. the total
long-range quark chromomagnetic moment equals to zero, which is in accord
with the flux tube \cite{buch} and minimal area \cite{bcp,bv} models.
One can see from Eq.~(\ref{vsd}) that for the spin-dependent part of
the potential this conjecture is equivalent to
the assumption about the scalar structure of confinement interaction
\cite{scal}. 
 
\section{Heavy quarkonium mass spectra}
Now we can calculate the mass spectra of heavy quarkonia with the account
of all relativistic corrections (including retardation effects) of order $v^2/c^2$
and one-loop radiative corrections. For this purpose we substitute the quasipotential
which is a sum of the spin-independent  (\ref{sipot}) and spin-dependent (\ref{vsd})
parts into the quasipotential equation (\ref{quas}). Then we multiply the resulting
expression from the left by the quasipotential wave function of a bound state and
integrate with respect to the relative momentum. Taking into account the accuracy
of the calculations, we can use for the resulting matrix elements the wave functions of
Eq.~(\ref{quas}) with the static potential \footnote{This static potential includes
also some  radiative corrections \cite{ty}.  The remaining radiative correction
term with logarithm in (\ref{sipot}), also not vanishing in the static limit, is treated
perturbatively.}  
\begin{equation}
V_{\rm NR}(r)=-\frac43\frac{\bar \alpha_V(\mu^2)}{r} +Ar+B.
\end{equation}
As a result we obtain the mass formula ($m_a=m_b=m$)
\begin{equation}
\label{mform}
\frac{b^2(M)}{2\mu_R}=W+\langle a\rangle\langle{\bf L}\cdot{\bf S}\rangle
+\langle b\rangle \langle\left[\frac{3}{r^2}
({\bf S}_a\cdot {\bf r})
({\bf S}_b\cdot {\bf r})-({\bf S}_a\cdot {\bf S}_b)\right] \rangle
+\langle c\rangle \langle{\bf S}_a\cdot {\bf S}_b\rangle,
\end{equation}
where
\begin{eqnarray}
W&=&\langle V_{\rm SI}\rangle+\frac{\langle {\bf p}^2\rangle}{2\mu_R},\cr
\langle{\bf L}\cdot{\bf S}\rangle&=& \frac12(J(J+1)-L(L+1)-S(S+1)),\cr
\langle\left[\frac{3}{r^2}
({\bf S}_a\cdot {\bf r})
({\bf S}_b\cdot {\bf r})-({\bf S}_a\cdot {\bf S}_b)\right] \rangle&=&
-\frac{6(\langle{\bf L}\cdot{\bf S}\rangle)^2+3\langle{\bf L}\cdot{\bf S}\rangle
-2S(S+1)L(L+1)}{2(2L-1)(2L+3)},\cr
\langle{\bf S}_a\cdot {\bf S}_b\rangle&=&\frac12\left(S(S+1)-\frac32\right),
\qquad
{\bf S}={\bf S}_a+{\bf S}_b,\nonumber
\end{eqnarray}
and $\langle a\rangle$, $\langle b\rangle$, $\langle c\rangle$ are the appropriate
averages over radial wave functions of Eqs.~(\ref{a})-(\ref{c}).  We use the usual
notations for heavy quarkonia classification: $n^{2S+1}L_J$, where $n$ is a radial
quantum number, $L$ is the angular momentum, $S=0,1$ 
is the total spin, and $J=L-S, L, L+S$  is  the total angular
momentum (${\bf J}={\bf L}+{\bf S}$).

The first term on the right-hand side of the mass formula (\ref{mform}) contains
all spin-independent contributions, the second term describes the spin-orbit
interaction, the third term is responsible for the tensor interaction, while the
last term gives the spin-spin interaction.

To proceed further we need to discuss the parameters of our model. There is the
following set of parameters: the quark masses ($m_b$ and $m_c$), the QCD
constant $\Lambda$ and renormalization point $\mu$ (see Eqs.~(\ref{alpha}), 
(\ref{sipot}), (\ref{vsd})) in the short-range part of the $Q\bar Q$ potential,
the slope $A$ and intercept $B$ of the
linear confining potential (\ref{nr}), the mixing
coefficient $\varepsilon$ (\ref{vlin}), the long-range anomalous chromomagnetic
moment $\kappa$ of the quark (\ref{kappa}), and the mixing parameter $\lambda_S$
in the retardation correction for the scalar confining potential 
(\ref{cret}).  As it was
already discussed
in Sec.~II, we can fix the values of the parameters $\varepsilon=-1$ and
$\kappa=-1$ from the consideration of radiative decays \cite{gf} and comparison
of the heavy quark expansion in our model \cite{fg,exc}
with the predictions of the heavy quark effective theory. We  fix the slope of
the linear confining potential $A=0.18$~GeV$^2$ which is a rather adopted
value. In order to reduce the number of independent parameters we assume
that the renormalization scale $\mu$ in the strong 
coupling constant $\alpha_s(\mu^2)$
is equal to the quark mass.~\footnote{Our numerical analysis 
showed that this is a good approximation, since the variation of $\mu$ does not increase 
considerably the quality of the mass spectrum fit.} We also varied the quark masses
in a reasonable range for the constituent quark masses. The numerical analysis
and comparison with experimental data
lead to the following values of our model parameters:
\begin{eqnarray}
&&m_c=1.55~{\rm GeV}, \quad m_b=4.88~{\rm GeV}, 
\quad \Lambda=0.178~{\rm GeV},\cr
&&A=0.18~{\rm GeV}^2, \quad B=-0.16~{\rm GeV},\quad \mu=m_Q~\ (Q=c,b), \cr
&&\varepsilon=-1, \quad \kappa=-1, \quad \lambda_S=0. \nonumber
\end{eqnarray}
The quark masses $m_{c,b}$ have usual values for constituent quark models
and coincide with those chosen in our previous analysis \cite{mass}
(see Sec.~II). The above
value of the retardation parameter $\lambda_S$ for the
 scalar confining potential
coincides with the minimal area low and flux tube models \cite{bv}, with
lattice results \cite{bali} and Gromes suggestion \cite{gromes}. The found
value for the QCD parameter $\Lambda$ gives the following
values for the strong coupling constants $\alpha_s(m_c^2)\approx 0.32$ and
$\alpha_s(m_b^2)\approx 0.22$.  

The results of our numerical calculations of the mass spectra of charmonium and
bottomonium are presented in Tables~\ref{charm} and \ref{bottom}. We see that
the calculated masses agree with experimental 
values within few MeV and this difference
is compatible with the estimates of the higher order corrections in $v^2/c^2$ and
$\alpha_s$. The model reproduces correctly both the positions of the centres of
gravity of the levels and their fine and hyperfine splitting.
Note that the good    
agreement of the calculated mass spectra with experimental data is achieved 
by systematic accounting for all relativistic corrections 
(including retardation corrections) of 
order $v^2/c^2$, both spin-dependent and 
spin-independent ones, while in most 
of potential models only the spin-dependent
corrections are included. 

\begin{figure}
\centerline{
\epsfysize=10cm
\epsfbox{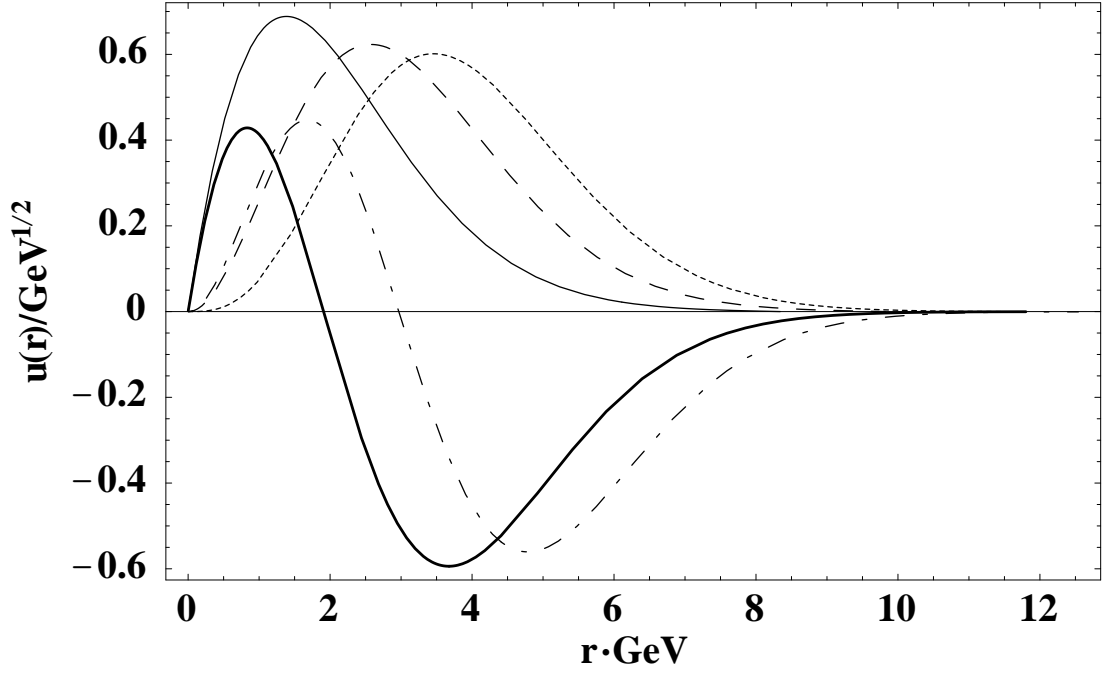}}
\caption{The reduced radial wave functions for charmonium. The solid line is for
$1S$, bold line for $2S$, long-dashed line for $1P$, dashed-dotted line for $2P$,
and dotted line for $1D$ states. }
\label{cm}
\end{figure}

\begin{figure}
\centerline{
\epsfysize=10cm
\epsfbox{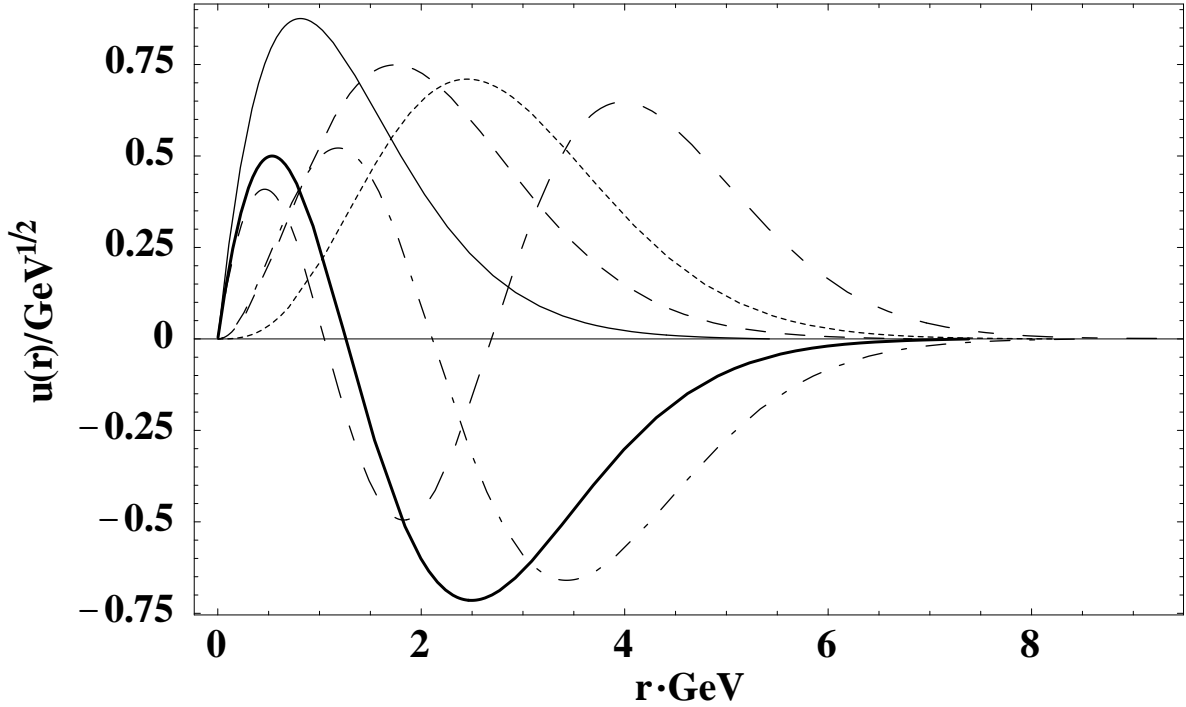}}

\caption{The same as in Fig.~1 for bottomonium and long-short-dashed line for
$3S$ state.}
\label{bt}
\end{figure}

The calculated mass spectra of charmonium and
bottomonium are  close to the results of
our previous calculation \cite{mass} where 
retardation effects in the confining potential and radiative corrections 
to the one-gluon exchange potential were not taken into account. 
Both calculations
give close values for the experimentally measured states as well as for
the yet unobserved ones. The inclusion of radiative corrections allowed to get
better results for the fine splittings of quarkonium states. 
Thus we can conclude
from this comparison that the
inclusion of retardation effects and spin-independent
one-loop radiative corrections resulted only in the slight 
shift ($\approx 10\%$)in the value of 
the QCD parameter $\Lambda$ and an approximately two-fold decrease of the 
constant $B$.~\footnote{Note that in Ref.~\cite{mass} we included this constant
both in vector and scalar parts, while  the present analysis indicates
 that the better
fit can be obtained if the constant $B$ is included only 
in the vector part (\ref{vlin}).} 
Such changes of parameters almost do not influence the wave functions. As
a result the decay matrix elements involving heavy quarkonium states remain
mostly unchanged.~\footnote{The changes in decay matrix elements are
of the same order of magnitude as the contributions of the 
higher order relativistic and 
radiative corrections.}  We plot the reduced radial wave 
functions $u(r)=rR(r)$ for
charmonium and bottomonium in Figs.~\ref{cm} and \ref{bt}.

\section{Conclusions}
In this paper we have considered the heavy quarkonium spectroscopy
in the framework of the relativistic quark model. Both relativistic
corrections of order $v^2/c^2$ and one-loop radiative corrections
to the short-range potential have been included into the calculation. Special
attention has been devoted to the role and the structure of retardation
corrections to the confining interaction. Our general analysis of
the retardation effects has shown that we have a good theoretical 
motivation to fix the form of retardation contributions to the vector 
potential in the form (\ref{vpr}) which corresponds to  the parameter
$\lambda_V=1$ in the generalized expression (\ref{cret}). On the contrary, 
the structure of the retardation contribution to the scalar potential
is less restricted from general analysis. This means that it is not possible
to fix the value of $\lambda_S$ in (\ref{cret}) on general grounds. Our
numerical analysis has shown that the value of $\lambda_S=0$ is
preferable. Thus for the energy transfer squared we have 
the symmetrized product  
(\ref{k0s}) for the vector potential and a half sum (\ref{k0hs}) for the scalar
potential, in agreement with lattice calculations \cite{bali} and minimal
area law and flux tube models \cite{bv}. The found structure of 
the spin-independent interaction (\ref{sipot}) with the
account of retardation contributions satisfies the BBP \cite{bbp}
relations (\ref{re}), which follow from the Lorentz invariance of the 
Wilson loop. 

In our calculations we have used the heavy quark-antiquark interaction
potential with the complete account of all relativistic corrections
of order $v^2/c^2$ and one-loop radiative corrections both for the
spin-independent and spin-dependent parts. The inclusion of these 
corrections allowed  to fit correctly the position of the centres 
of gravity of the heavy quarkonium levels as well as
their fine and hyperfine splittings. Moreover,
the account for radiative corrections results in a better description 
of level splittings. The values of the main parameters of our quark model
such as the slope of the confining linear potential $A=0.18$ GeV$^2$, 
the mixing coefficient $\varepsilon=-1$ of scalar and vector confining 
potentials and the long-range anomalous chromomagnetic quark
moment $\kappa=-1$ used in the present analysis are kept 
the same as they were fixed  from the
previous consideration of radiative decays \cite{gf} and the heavy
quark expansion \cite{fg,exc}. The value of $\varepsilon=-1$ 
implies that the confining 
quark-antiquark potential in heavy mesons has predominantly a Lorentz-vector
structure, while the scalar potential is anticonfining and helps to
reproduce the initial nonrelativistic potential. On the other hand,
the value of $\kappa=-1$ supports the conjecture that the long-range
confining forces are dominated by chromoelectric interaction and that 
the chromomagnetic interaction vanishes, which is in accord with the dual
superconductivity picture \cite{bbbpz} and flux tube model \cite{buch}. 
  
The presented results for the charmonium and bottomonium mass
spectra agree well with the available experimental data.
It is of great interest to consider the predictions for the masses of the $^1S_0$ and $D$
levels of bottomonium, which have not yet been observed experimentally. The difficulty
of their experimental observation is that these states (except $^3D_1$) cannot be
produced in $e^+e^-$ collisions, since their quantum numbers are not the same as the quantum numbers of the photon. Therefore, in search for these states one must 
investigate decay processes of vector ($^3S_1$) levels. We discussed the possibility 
of observation of these states in radiative decays in Ref.~\cite{gf}. Note that the small
value predicted for the hyperfine splitting $M(\Upsilon)-M(\eta_b)\cong 60$~MeV
leads to difficulties in observation of the $\eta_b$ state. 

Recently it was argued
\cite{bmb} that the account of relativistic kinematics 
substantially modifies the description of the charmonium fine structure
and, in particular, leads 
to considerably larger values of the $2^3P_J$ splittings 
than in the nonrelativistic
limit. Both our previous calculation \cite{mass} and the present one  
confirm this observation. Our prediction for the charmonium $2^3P_0$ mass
lies close to the
prediction of Ref.~\cite{bmb} and slightly lower than the
 $D\bar D^*$ threshold. However,
the fact that this state is above $D\bar D$  and close to $D\bar D^*$ thresholds makes
threshold effects very important and can considerably influence the quark model
prediction.

\acknowledgements
We thank A.M. Badalyan, G. Bali,
N. Brambilla, M.I. Polikarpov and V.I. Savrin
for useful discussions of the results. Two of us (R.N.F. and V.O.G.)
are grateful to the particle theory group of Humboldt University for
the kind hospitality. The work of R.N.F. and V.O.G. was supported in
part by the Deutsche Forschungsgemeinschaft under contract Eb 139/1-3.

\begin{table}
\caption{Charmonium mass spectrum. }
\label{charm}
\begin{tabular}{ccccc}
State ($n^{(2S+1)}L_J$)& Particle &  Theory & Experiment \cite{pdg}&
 Experiment \cite{bes} \\
\hline
$1^1S_0$& $\eta_c$ & 2.979 & 2.9798 &2.9758\\
$1^3S_1$& $J/\Psi$ & 3.096 & 3.09688 &\\
&&&&\\
$1^3P_0$& $\chi_{c0}$ & 3.424 & 3.4173 & 3.4141\\
$1^3P_1$& $\chi_{c1}$ & 3.510 & 3.51053 &\\
$1^3P_2$& $\chi_{c2}$ & 3.556 & 3.55617 &\\
&&&&\\
$2^1S_0$& $\eta_c'$ & 3.583 & 3.594 &\\
$2^3S_1$& $\Psi'$     & 3.686 & 3.686 &\\ 
&&&&\\
$1^3D_1$&  & 3.798 & $3.7699^*$ &\\
$1^3D_2$&  & 3.813 & &\\
$1^3D_3$&  & 3.815 & &\\
&&&&\\
$2^3P_0$& $\chi'_{c0}$ & 3.854 & &\\
$2^3P_1$& $\chi'_{c1}$ & 3.929 & &\\
$2^3P_2$& $\chi'_{c2}$ & 3.972 & &\\
&&&&\\
$3^1S_0$& $\eta_c''$ & 3.991 & &\\
$3^3S_1$& $\Psi''$     & 4.088 & $4.040^*$ &\\
&&&&\\
$2^3D_1$&  & 4.194 & $4.159^*$ &\\
$2^3D_2$&  & 4.215 & &\\
$2^3D_3$&  & 4.223 & &\\ 
\end{tabular}
$^*$ Mixture of $S$ and $D$ states 
\end{table} 

\begin{table}
\caption{Bottomonium mass spectrum. }
\label{bottom}
\begin{tabular}{ccccc}
State ($n^{(2S+1)}L_J$)& Particle &  Theory & Experiment \cite{pdg}
&Experiment \cite{cleo} \\
\hline
$1^1S_0$& $\eta_b$ & 9.400 &  &\\
$1^3S_1$& $\Upsilon$ & 9.460 & 9.46037 &\\
&&&&\\
$1^3P_0$& $\chi_{b0}$ & 9.864 & 9.8598 & 9.8630\\
$1^3P_1$& $\chi_{b1}$ & 9.892 & 9.8919 & 9.8945\\
$1^3P_2$& $\chi_{b2}$ & 9.912 & 9.9132 &  9.9125\\
&&&&\\
$2^1S_0$& $\eta_b'$ & 9.990 & &\\
$2^3S_1$& $\Upsilon'$  & 10.020 & 10.023 &\\ 
&&&&\\
$1^3D_1$&  & 10.151 &  & \\
$1^3D_2$&  & 10.157 &  &\\
$1^3D_3$&  & 10.160 & &\\
&&&&\\
$2^3P_0$& $\chi'_{b0}$ & 10.232 & 10.232 & \\
$2^3P_1$& $\chi'_{b1}$ & 10.253 & 10.2552 & \\
$2^3P_2$& $\chi'_{b2}$ & 10.267 & 10.2685 & \\
&&&&\\
$3^1S_0$& $\eta_b''$ & 10.328 & & \\
$3^3S_1$& $\Upsilon''$ & 10.355 & 10.3553 &\\
&&&&\\
$2^3D_1$&  & 10.441 &   & \\
$2^3D_2$&  & 10.446 &  &\\
$2^3D_3$&  & 10.450 &  &\\ 
&&&&\\
$3^3P_0$& $\chi''_{b0}$ & 10.498 &  & \\
$3^3P_1$& $\chi''_{b1}$ & 10.516 &  & \\
$3^3P_2$& $\chi''_{b2}$ & 10.529 &  & \\
&&&&\\
$4^1S_0$& $\eta_b'''$ & 10.578 & & \\
$4^3S_1$& $\Upsilon'''$ & 10.604 & 10.580 &\\
\end{tabular}
\end{table}

\end{document}